\documentstyle[epsf,aps,prb]{revtex}

\begin{document}

\twocolumn[\hsize\textwidth\columnwidth\hsize\csname
@twocolumnfalse\endcsname

\title{The Heisenberg model on the 1/5-depleted square lattice \\ 
and the CaV$_4$O$_9$ compound}

\author{ L. O. Manuel, M. I. Micheletti, A. E. Trumper,
 and H. A. Ceccatto}
\address{Instituto de F\'{\i}sica Rosario, Consejo Nacional de Investigaciones
Cient\'{\i}ficas y T\'ecnicas \\ and Universidad Nacional de Rosario, Bvd. 27
de Febrero 210 Bis, 2000 Rosario, Rep\'ublica Argentina }
\maketitle
\maketitle

\begin{abstract}
We investigate the ground state structure of the Heisenberg model
on the 1/5-depleted square lattice for arbitrary values of the 
first- and second-neighbor exchange couplings. By using a mean-field
Schwinger-boson approach we present a unified description of the
rich ground-state phase diagram, which includes the plaquette and dimer
resonant-valence-bond phases, long- and short-range N\'eel orders,
an incommensurate phase and other magnetic orders with complex magnetic
unit cells. We also discuss some implications of our results for the 
experimental realization of this model in the CaV$_4$O$_9$ compound.
\end{abstract}

]

\narrowtext

The CaV$_{4}$O$_{9}$ is the first example of a quasi two dimensional
magnetic system with a spin gap.\cite{tani} As such, it has received much
attention recently, with a large part of this activity devoted to
understanding the mechanism of the gap formation.\cite{katoh1}$^{-}$
\cite{fuku} 
Most of these works have considered the Heisenberg model on the so called CAVO
lattice, {\it i.e}., the 1/5-depleted square lattice of Fig. 1, with both
first- and second-neighbor interactions (Following the notation in Ref. 3, 
we will call $J_{1}$ and $J_{1}^{\prime}$ to the nearest-neighbor
intra and interplaquette couplings respectively, and, accordingly, $J_{2}$
and $J_{2}^{\prime}$ for the second-neighbor interactions). 

\begin{figure}[ht]
\begin{center}
\epsfysize=4.0cm
\leavevmode
\epsffile{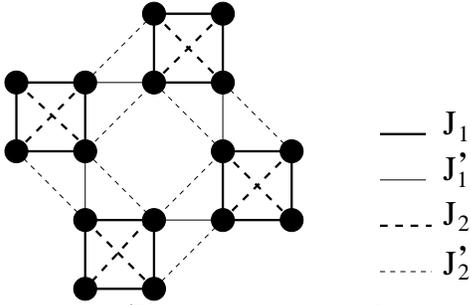}
\caption{ The 1/5-depleted square lattice, also called a CAVO lattice. We
indicate also the first- and second-neighbor exchange couplings between
vanadium atoms in the CaV$_4$O$_9$ compound.}
\end{center}
\end{figure}

A variety of
methods, including quantum Monte Carlo,\cite{katoh1,troyer} perturbative and
high-temperature series expan\-sions, \cite{gelfa1} exact diagonalization,\cite
{sano} different mean-field approximations, \cite{ueda,albre1,stary},
renormalization group methods\cite{white}, and other techniques\cite{sach,miya},
have been applied 
to this model, leading to the conviction that it has a spin gap for 
$J_{1}\gtrsim
J_{1}^{\prime}$ and/or large enough frustrating second-neighbor
interactions. Based on simple considerations, most authors have assumed $%
J_{1}^{\prime}\approx J_{1}$, $J_{2}^{\prime}\approx J_{2}$
and $J_{2}\approx 0.5J_{1}$, and, to lowest order, identified the
gap as the singlet-triplet gap above the plaquette resonant-valence bond
(PRVB) state obtained from the four nearest-neighbor V atoms coupled by $%
J_{1}$.\cite{ueda} However, by using LDA calculations Pickett\cite{picke}
concluded that the second-neighbor interactions might be dominant, with $%
J_{2}\approx 2J_{1}$. In this case the singlet ground state would
correspond to weakly-coupled metaplaquettes formed by V atoms connected by
the $J_{2}^{\prime}$ bonds. On the other hand, from the analysis of neutron
inelastic scattering data Kodama {\it et al.}\cite{koda} estimated $%
J_{2}\approx 0.1J_{2}^{\prime}$, $J_{1}^{\prime}\approx
J_{1}\approx 0.4J_{2}^{\prime}$ in order to reproduce the dispersion
of the lowest triplet excitation. Moreover, other authors have suggested
that the exchange couplings could be temperature dependent due to
spin-phonon interactions.\cite{stary}

All the above mentioned works assumed that a single non-degenerate orbital
on V atoms is occupied. Marini and Khomskii\cite{marini} argued, however,
that orbital ordering and associated structural distortions determine the
magnetic properties of the vanadates CaV$_n$O$_{2n+1}$, and proposed a
structure of weakly-coupled dimers along $J_1^{\prime}$ bonds for the
ground state of CaV$_4$O$_9$. Recent work by Gelfand and Singh\cite{gelfa2}
concluded that this proposal is not consistent with experimental
observations, and that the PRVB and metaplaquettes scenarios are both able
to produce a reasonable explanation for the observed gap and uniform
susceptibility. In a very recent work,\cite{katoh2} Katoh and Imada pointed
out serious limitations of the simple single-orbital model to reproduce
results from neutron inelastic scattering. They partially solved these
discrepancies between theoretical predictions and experimental observations
by considering the effects of orbital degeneracy and orbital order, which
lead to an effective spin Hamiltonian with exchange couplings strongly
dependent on patterns of orbital occupancy.

The above brief summary of previous work on CaV$_4$O$_9$ points to the
difficulties found to determine the real values of the competing couplings
involved. Consequently, it seems worth to us the investigation of the
general ground-state structure of the quantum Heisenberg antiferromagnet on
the CAVO lattice for arbitrary values of $J_1$, $J_1^{\prime}$ and $J_2$
(we take $J_2^{\prime}=J_2$ for simplicity), in order to identify
possible phases with the properties observed experimentally. We report here
the results of such an investigation, which is based on the use of the
mean-field Schwinger boson approximation.

\begin{figure}[ht]
\begin{center}
\epsfysize=6.5cm
\leavevmode
\epsffile{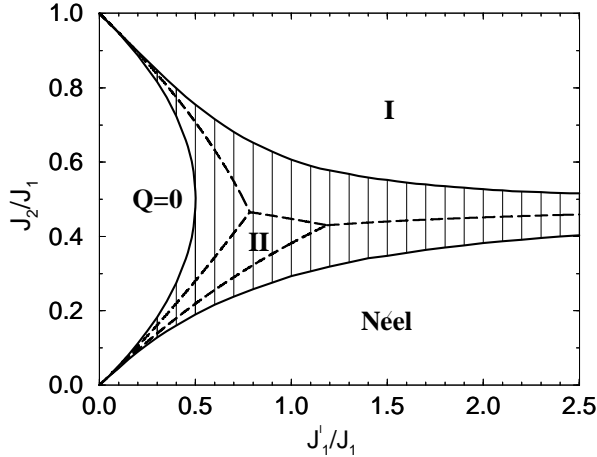}
\caption{Ground-state structure of the classical Heisenberg model on the CAVO
lattice. Full lines indicate the stability regions of phases I, N\'{e}el and 
$Q=0$. In the shaded area an incommensurate phase is stabilized in the
thermodynamic limit. The dashed lines indicates the stable phases I, II,
N\'{e}el and $Q=0$ for a cluster of 32 spins.}
\end{center}
\end{figure}

Let us start by considering first the ground state of the classical
Heisenberg model on the lattice of Fig. 1. A standard analysis\cite{fuku}
shows that even in this case the model turns out to be very rich, as shown
in Fig. 2. The stability regions of the different phases are indicated by
full lines; we also show, for further use, the stable phases in lattices of
up to 32 spins (dashed lines). The $Q=0$ and N\'{e}el phases have
antiferromagnetic order in the plaquettes which repeats itself along the
translation vectors ${\bf \delta}_{1}=(2,1)a$, ${\bf \delta}_{2}=(-1,2)a$
with magnetic wavevectors ${\bf Q}=(0,0)$ and ${\bf Q}=(\pi ,\pi )$
respectively ($a$ is the nearest neighbor V-V distance). Phases I and II
have both a 8-spin magnetic unit cell, translation vectors ${\bf \delta }%
_{1}=(1,3)a$, ${\bf \delta }_{2}=(-3,1)a$ and a magnetic wavevector ${\bf Q}%
=(\pi ,\pi )$, with the corresponding spin arrangements in the unit cells
shown in Fig. 3a,b. The angle $\varphi $ verifies $\tan \varphi
=-J_{1}^{\prime}/2J_{2}^{\prime}$ in phase I and $\tan \varphi
=-2J_{1}/J_{1}^{\prime}$ in phase II. Finally, the shaded area is the
stability region of an incommensurate phase very difficult to characterize.
This makes the consideration of the quantum nature of the spins quite
involved in this phase (see below).

\begin{figure}[ht]
\begin{center}
\epsfysize=4.0cm
\leavevmode
\epsffile{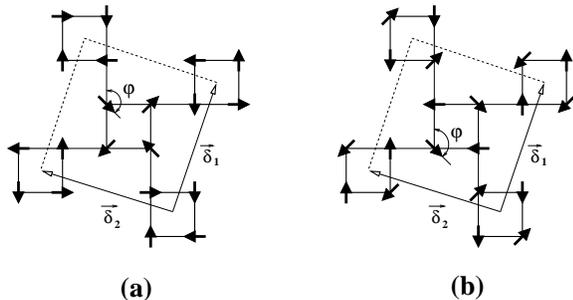}
\caption{Magnetic unit cells and displacement vectors ${\bf \delta}_{1}$ and 
${\bf \delta}_{2}$ corresponding to a) Phase I and b) Phase II (see main text).} 
\end{center}
\end{figure}

We must now correct the classical diagram of Fig. 2 by introducing the
effects of quantum fluctuations. In order to do so, we will use the
mean-field Schwinger boson theory\cite{auer}, which has already been applied
to the CAVO lattice without frustrating second-neighbor interactions.\cite
{albre1} This approach is suitable for the problem at hand since it allows
to treat magnetic and non-magnetic quantum phases in an unified way.
Moreover, despite the fact that the Schwinger boson theory is known to
overestimate the stability of magnetic phases, it most often provides a
qualitatively reliable phase structure. On the other hand, its predictions
can be systematically corrected by the inclusion of fluctuations in the
mean-field order parameters,\cite{tmgc} which has been shown to bring
Schwinger boson results well in line with exact values on finite lattices
and with the predictions of other more accurate methods in the thermodynamic
limit.

Using the standard representation of spin operators in terms of two boson
fields linked by the local boson number restrictions, quantum corrections to
the phase diagram of Fig. 2 can be obtained by simply choosing the right
initial structure of the order parameters in solving the self-consistent
equations.\cite{cgt} Although this is a matter of a straightforward energy
minimization, the large number of spins in the magnetic Brillouin zone makes
the calculations involved and computationally demanding. In addition, since
we have no simple characterization of the incommensurate phase, in this
region we had to rely on an extensive search for global solutions of the
minimization equations. The strategy used was first to determine the
classical ground state in clusters of up to 64 spins, and then searching for
the absolute energy minimum in the quantum incommensurate phase starting
from these large magnetic cells. In particular, up to 32 spins the stable
classical phases are those showed in Fig. 2 with dashed lines. In addition
to the commensurate phases above described, there is another phase stable
only along the dashed line that separates phases I and II; it has a 4-spin
magnetic cell and magnetic wavevector ${\bf Q}=(0,\pi)$. This classical
phase is particularly interesting as a starting point for the search of
order parameters in the quantum incommensurate phase.
This is so because no pair of classical vectors in the magnetic motif are
collinear or anticollinear, which means that the ferromagnetic and
antiferromagnetic channels are both active in the minimization. Furthermore,
for clusters with more than 32 spins the resulting classical stable phases
are incommensurate (span the cluster). We stress that this investigation of
the incommensurate phase is allowed only because of the particular
properties of the Schwinger boson theory, since other semiclassical
approaches (spin wave theory, for instance) require the precise knowledge of
the classical ground-state structure. Moreover, the inclusion of quantum
(non-magnetic) phases in the diagram of Fig. 2 is possible because the
Schwinger bosons preserve the rotational invariance of the Heisenberg
Hamiltonian when there is no condensation (magnetization) in the system.

\begin{figure}[ht]
\begin{center}
\epsfysize=6.5cm
\leavevmode
\epsffile{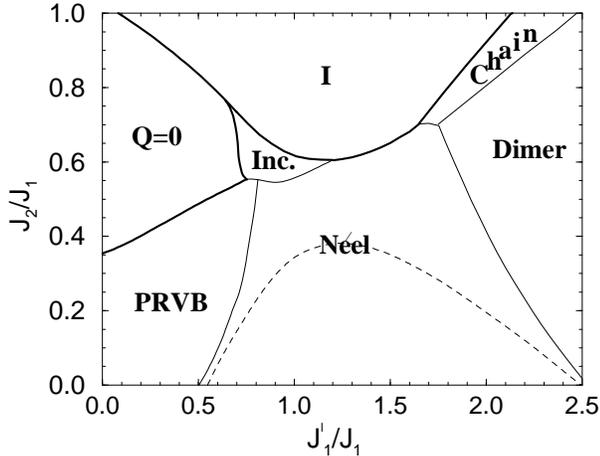}
\caption{Ground-state structure of the quantum Heisenberg model on the CAVO
lattice. Full thick(thin) lines indicate first(second)-order phase
transitions; the dashed line separates long-range and short-range N\'{e}el
orders. See main text for a description of the different phases.}
\end{center}
\end{figure}

After the consideration of quantum fluctuations at mean-field order, the
ground state has the complex structure shown in Fig. 4. The $Q=0$-PRVB 
transition is weak first order, while the
incommensurate-short-range N\'{e}el transition might be a very weak one
instead of the second-order transition indicated. The Chain phase
corresponds to disconnected zig-zag chains along $J_{1}^{\prime}$%
and $J_{2}$ bonds, with short-range antiferromagnetic correlations between
nearest-neighbor spins in the chains (this phase is essentially the ground
state proposed by Marini and Khomskii\cite{marini}). The PRVB and dimer
phases are the standard singlet phases discussed in previous works,
consisting respectively of isolated plaquettes made up by $J_{1}$ bonds and
the $J_{1}^{\prime}$ dimer covering. For $J_{2}^{\prime}\gg J_{1}^{\prime}
$ phase I goes to Pickett's weakly-coupled metaplaquette scenario.\cite
{picke} Notice that in this limit the angle $\varphi $ between
metaplaquettes becomes $\pi /2$, instead of $0$ or $\pi $ expected from
order-from-disorder ideas. Of course, this phase diagram might be strongly
modified by order-parameter fluctuations, but it gives an idea of the model
complexity and becomes a starting point for these much more involved
calculations. 

Notice that the most studied region in connection with the CaV$_{4}$O$_{9}$
compound ($J_{1}^{\prime}\approx J_{1}$, $J_{2}\approx 0.5J_{1}$)
corresponds to a region where several distinct phases merge. On the other
hand, Pickett's proposal $J_{2}\approx 2J_{1}$ \cite{picke} is well
inside the phase I. Then, it is of interest to discuss in more detail what
happens along the line $J_{1}^{\prime}=J_{1}$ and arbitrary $J_{2}$. In
Fig. 5 we see the behavior of the energy, with the continuous transition
from the short-range N\'eel order to the incommensurate phase at $%
J_{2}\simeq 0.55J_{1}$, and then a first-order transition from this last
phase to phase I at $J_{2}\simeq 0.62J_{1}$. Fig. 6 shows the corresponding
magnetization curve, which indicates a robust magnetic order in phase I and
a rather peculiar behavior in the incommensurate phase. In this last phase
the magnetization $m$ has an abrupt drop to zero near the
transition to the short-range N\'eel order, which is in agreement with the
smooth behavior shown by the energy. However, the incommensurate magnetic
vector seems to have a discontinuity before reaching ($\pi,\pi$) (see Fig.
7). Nevertheless, this could be only a problem with the numerical accuracy,
since the convergence of the selfconsistent equations becomes more and more
difficult near the transition. Another remarkable characteristic of the 
magnetization curve is the local minimum in the incommensurate phase. Since 
the behavior of $%
{\bf Q}$ is smooth across these points, this could indicate that the spins
inside the magnetic unit cell rearrange without changing the incommensurate
wavevector. Such an explanation is supported by the small discontinuities we
observed in the short-range correlations on both sides near the deep local
minimum of $m$. Finally, in Fig. 6 the dashed and full lines in the 
incommensurate phase give the magnetization of different
sublattices. These two sublattices
correspond to the zigzagging paths along $J_{1}^{\prime}$ and $%
J_{2}$ bonds, in one direction and in the direction perpendicular to it
respectively. However, in this case the paths are not disconnected like in
the Chain phase, since correlations in $J_{1}$ bonds do not
vanish. In passing, we note that to obtain this solution from the
selfconsistent equations one must allow for different values of the Lagrange
multipliers that impose the boson-number restriction on inequivalent sites
of the magnetic unit cell.

\begin{figure}[ht]
\begin{center}
\epsfysize=6.5cm
\leavevmode
\epsffile{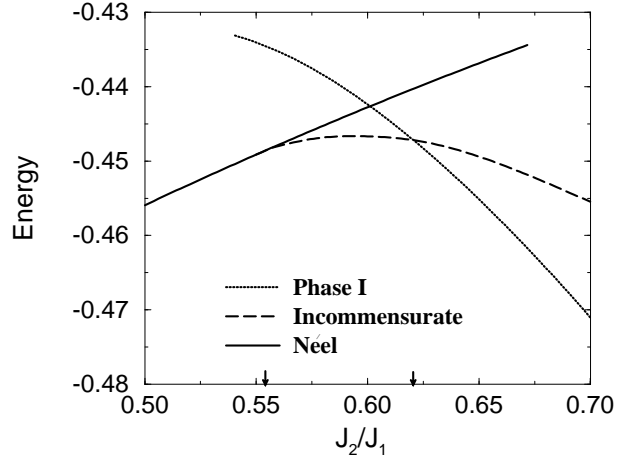}
\caption{Ground-state energy along the line $J_{1}^{\prime }=J_{1}$ in
parameter space. The arrows at $J_{2}/J_{1}\simeq 0.55$ and $0.62$ indicate
the points where there are phase transitions.}
\end{center}
\end{figure}

\begin{figure}[ht]
\begin{center}
\epsfysize=6.5cm
\leavevmode
\epsffile{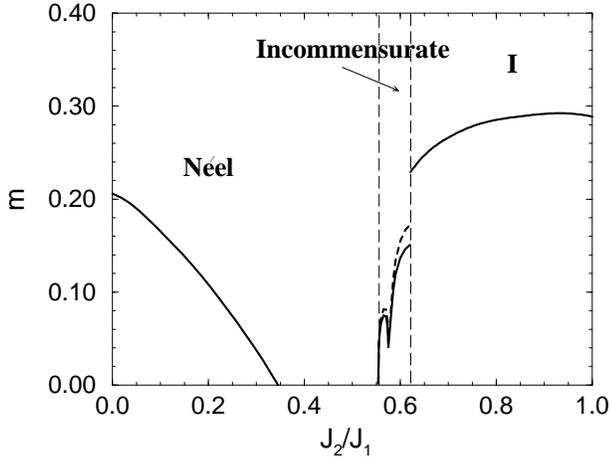}
\caption{Magnetization $m$ as a function of $J_{2}/J_{1}$ for $J_{1}^{\prime
}=J_{1}.$ The full and dashed lines in the incommensurate phase correspond
to the magnetization of two different sublattices.}
\end{center}
\end{figure}

\begin{figure}[ht]
\begin{center}
\epsfysize=6.0cm
\leavevmode
\epsffile{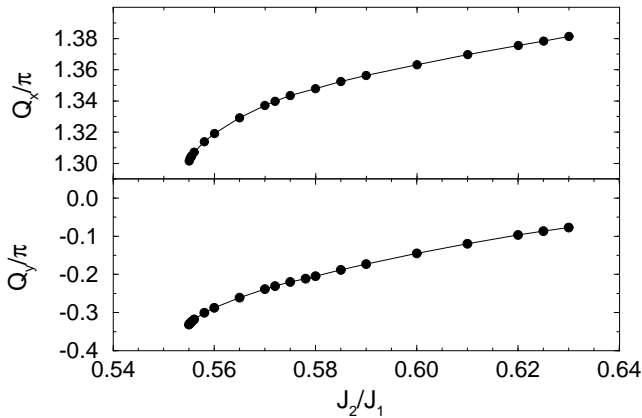}
\caption{The magnetic wavevector ${\bf Q}/\pi $ in the incommensurate phase
as a function of $J_{2}/J_{1}$ for $J_{1}^{\prime }=J_{1}$.}
\end{center}
\end{figure}

In conclusion, we have determined the general ground-state phase structure
of the frustrated Heisenberg model on the CAVO lattice by means of the
Schwinger boson mean-field theory. In particular, near the most studied
parameter region for the CaV$_{4}$O$_{9}$ compound we
identified several competing phases. Although the relative stability of
these phases can be strongly modified by the introduction of Gaussian
fluctuations in the order parameters, it is nonetheless quite satisfactory
to have a unified picture of the phase diagram at mean-field order.
Moreover, it provides a necessary starting point for the more involved
one-loop calculations. On the other hand, despite the fact that some of the
phases have long-range order --and, consequently, no gap--, the reduced
magnetization values suggest that fluctuation corrections might easily
destroy the magnetic order and produce the gapped phase observed
experimentally (like what happens, for instance, in the $J_{1}-J_{2}$ model
on the square lattice\cite{tmgc}). In the absence of frustration, the
wavenumber of the lowest energy excitation (corresponding to the
singlet-triplet gap in the PRVB scenario) has been predicted from
perturbation expansions\cite{gelfa1} and variational Monte Carlo\cite{miya}
to be ($\pi ,\pi $), and it shifts to incommensurate wavenumbers when $J_{2}$
becomes relatively large. Our results for the N\'{e}el, PRVB and I phases
indicate also a ($\pi ,\pi $) wavector for the lowest excitations, while the
incommensurate phase has a ${\bf Q}$ that changes with $J_{2}$ as shown in
Fig. 7. However, recent neutron inelastic scattering data for CaV$_{4}$O$_{9}
$\cite{koda} are consistent with ${\bf Q}=(0,0)$, like in a
disordered $Q=0$ phase. Finally, we are currently exploring the
possibility of introducing fluctuations in the order parameters, although
these are quite involved calculations.


\begin{thebibliography}{99}

\bibitem{tani} S. Taniguchi {\it et al.}, J. Phys. Soc. Jpn. {\bf 64}, 2758 
(1995).

\bibitem{katoh1} N. Katoh and M. Imada, J. Phys. Soc. Jpn. {\bf 64}, 4105 
(1995).

\bibitem{gelfa1} M. P. Gelfand {\it et al.}, Phys. Rev. Lett. {\bf 77}, 2794 
(1996).

\bibitem{troyer} M. Troyer, H. Kontani and K. Ueda, Phys. Rev. Lett. {\bf 76}, 
3822 (1996).

\bibitem{sano} K. Sano and K. Takano, J. Phys. Soc. Jpn. {\bf 65}, 46 (1996);
 M. Albrecht, F. Mila and D. Poilblanc, Phys. Rev. B {\bf 54},
15 856 (1996).

\bibitem{ueda} K. Ueda, H. Kontani, M. Sigrist, and P. A. Lee, Phys. Rev. Lett.
{\bf 76}, 1932 (1996).

\bibitem{albre1} M. Albrecht and F. Mila, Phys. Rev. B {\bf 53}, R2945 (1996).

\bibitem{stary}  O. A. Starykh {\it et al.}, Phys. Rev. Lett. {\bf 77}, 2558 
(1996).

\bibitem{white} S. R. White, Phys. Rev. Lett. {\bf 77}, 3633 (1996).

\bibitem{sach} S. Sachdev and N. Read, Phys. Rev. Lett. {\bf 77}, 4800 (1996).

\bibitem{miya} T. Miyazaki and D. Yoshioka, J. Phys. Soc. Jpn. {\bf 65}, 2370
 (1996).

\bibitem{picke} W. E. Pickett, preprint cond-mat/9704203.

\bibitem{koda} K. Kodama {\it et al.}, J. Phys. Soc. Jpn. {\bf 65}, 1941 
(1996); K. Kodama {\it et al.}, {\it ibid.}, {\bf 66}, 793 (1997).

\bibitem{marini} S. Marini and D. I. Khomskii, preprint cond-mat/9703130.

\bibitem{gelfa2} M. P. Gelfand and R. R. P. Singh, preprint cond-mat/9705122.

\bibitem{katoh2} N. Katoh and M. Imada, preprint cond-mat/9712051.

\bibitem{fuku} Y. Fukumoto, J. Phys. Soc. Jpn. {\bf 66}, 2178 (1997).

\bibitem{auer} A. Auerbach and D. Arovas, Phys. Rev. Lett. {\bf 61}, 617
(1988); D. Arovas and A. Auerbach, Phys. Rev. B {\bf 38}, 316 (1988).

\bibitem{tmgc} A. E. Trumper, L. O. Manuel, C. J. Gazza and H. A. Ceccatto, 
Phys. Rev. Lett. {\bf 78}, 2216 (1997). 

\bibitem{cgt} H. A. Ceccatto, C. J. Gazza and A. E. Trumper, Phys. Rev. B
{\bf 47}, 12 329 (1993). 

\end{thebibliography}
\end{document}